\documentclass[aps,pra,twocolumn]{revtex4-2}
\usepackage{epsfig,amssymb, amsmath,fancyhdr,upgreek,color}
\newcommand{\bem}{{\bf e}_m}
\newcommand{\bemp}{{\bf e}_{m'}}
\newcommand{\betm}{{\bf e}_{\tilde{m}}}
\newcommand{\bfN}{{\bf N}}
\newcommand{\tbfN}{{\tilde{\bf N}}}
\newcommand{\bzeta}{{\boldsymbol \zeta}}
\newcommand{\bhata}{{\hat{\boldsymbol a}}}

\begin{document}
\title{Performance analysis for high-dimensional Bell-state quantum illumination}
\author{Jeffrey H. Shapiro}
\email{jhs@mit.edu}
\affiliation{Research Laboratory of Electronics, Massachusetts Institute of Technology, Cambridge, Massachusetts 02139 USA}
\date{September 19, 2024}

\begin{abstract}
Quantum illumination (QI) is an entanglement-based protocol for improving lidar/radar detection of unresolved targets beyond what a classical lidar/radar of the same average transmitted energy can do.  Originally proposed by Lloyd as a discrete-variable quantum lidar, it was soon shown that his proposal offered no quantum advantage over its best classical competitor.  Continuous-variable, specifically Gaussian-state, QI has been shown to offer true quantum advantage, both in theory and in table-top experiments.  Moreover, despite its considerable drawbacks, the microwave version of Gaussian-state QI continues to attract research attention.  Recently, however, Pannu~\emph{et al}.~(arXiv:2407.08005 [quant-ph]) have: (1) combined the entangled state from Lloyd's QI with the channel models from Gaussian-state QI; (2) proposed a new positive operator-valued measurement for that composite setup; and (3) claimed that, unlike Gaussian-state QI, their QI achieves the Nair-Gu lower bound on QI target-detection error probability at all noise brightnesses.  Pannu~\emph{et al}.'s analysis was asymptotic, i.e., it presumed infinite-dimensional entanglement. This paper works out the finite-dimensional performance of Pannu~\emph{et al}.'s QI.  It shows that there is a threshold value for the entangled-state dimensionality below which there is no quantum advantage, and above which the Nair-Gu bound is approached asymptotically.  Moreover, with both systems operating with error-probability exponents 1\,dB lower than the Nair-Gu bound's, Pannu~\emph{et al}.'s QI requires much \emph{higher} entangled-state dimensionality than does Gaussian-state QI to achieve useful error probabilities in both high-brightness (100\,photons/mode) and moderate-brightness (1\,photon/mode) noise.  Furthermore, neither system has appreciable quantum advantage in low-brightness ($\ll1$\,photon/mode) noise.  
\end{abstract}

\maketitle

\section{Introduction \label{Intro}}
Lidar (at optical wavelengths) and radar (at microwave wavelengths) transmit electromagnetic radiation into a region of interest to discern characteristics of objects, which may or may not be present therein, based on the return radiation collected from that region~\cite{Skolnik2002}.  Despite electromagnetic radiation being fundamentally quantum mechanical~\cite{Louisell1973}, it is only recently that the use of quantum resources, specifically entanglement, has been considered for improving on the performance of classical lidars or radars, i.e., those whose performance can be correctly assessed \emph{without} treating their radiation in quantum terms.  Of special note in regard to quantum lidar or radar is quantum illumination (QI), in which entangled signal and idler beams are created, with the signal transmitted into the region of interest, while the idler is retained for a joint measurement with the returned radiation, see Refs.~\cite{Shapiro2020,Sorelli2022} for reviews of QI.  Inasmuch as QI target detection is the present paper's focus, it behooves us to briefly delve into some relevant history.

Lloyd~\cite{Lloyd2008} coined the term ``quantum illumination'' for a lidar that transmitted a sequence of $M$-mode single-photon states while retaining their maximally-entangled single-photon companions for a joint measurement with the returned radiation.  He assumed that the environment being probed never returned more than one photon in response to each transmission, and that returned photon was either background noise or a target reflection.  Lloyd compared his QI performance with that of a single-photon (SP)  lidar that probed the environment with the same state as his QI lidar, but had no stored idler.  He argued that QI's entanglement would make it difficult for a background photon to masquerade as the entangled companion of QI's stored idler.  Indeed, as compared to SP target detection, Lloyd's QI target detection afforded a factor-of-$M$ improvement in error-probability exponent in his high-noise regime, i.e., when it is vastly more probable that the returned radiation from a single transmission is due to background as opposed to target reflection.  That said, Lloyd's QI and SP lidars are \emph{both} quantum lidars, as they employ nonclassical transmitter states.  Thus, when Shapiro and Lloyd~\cite{Shapiro2009} compared Lloyd's QI with its best classical-lidar counterpart it turned out that the former could do no better than the latter, and could perform much worse.  As a result, interest in Lloyd's discrete-variable QI languished and was supplanted by interest in a continuous-variable version of QI, viz., Tan~\emph{et al}.'s Gaussian-state QI lidar~\cite{Tan2008}.  

In Gaussian-state QI, $M$-mode pulses of quadrature-entangled signal and idler are produced, with the former probing the region of interest and the latter stored for a joint measurement with the returned radiation.  For detecting the possible presence of a weakly-reflecting target embedded in high-brightness ($\gg$1 photon/mode) background radiation, Tan~\emph{et al}.\@ showed that Gaussian-state QI offered a 6\,dB advantage in error-probability exponent over its best classical competitor of the same transmitted energy.  Remarkably, this 6\,dB performance advantage is obtained \emph{only} in high-brightness noise, where the initial entanglement is destroyed, not in low-brightness ($\ll$1 photon/mode) noise such as exists at optical wavelengths~\cite{Shapiro2005}.  Consequently, intense interest in Gaussian-state QI did not develop until Barzanjeh~\emph{et al}.~\cite{Barzanjeh2015} showed how it could be used at microwave wavelengths, where weak target returns and high-brightness noise are the norm.

Initial table-top Gaussian-state QI experiments using sub-optimum receiver architectures have been reported for the optical region (with artificially-injected high-brightness noise)~\cite{Zhang2015} and for the microwave~\cite{Assouly2023}.  Although these  experiments only demonstrated $\sim$20\% signal-to-noise ratio gains over their best classical competitors, they did verify Gaussian-state QI's unique capability of providing an entanglement-based quantum advantage in an entanglement-breaking target-detection scenario.  As explained in Refs.~\cite{Shapiro2020,Sorelli2022}, Gaussian-state QI in the microwave faces enormous hurdles before its target-detection advantage can find a realistic use case.  These include:  Gaussian-state QI's need to interrogate one resolution bin at a time; Gaussian-state QI's need for a quantum memory to store its high time-bandwidth product    idler; Gaussian-state QI's requiring radiation with likely-to-be unattainably high time-bandwidth products; and Gaussian-state QI's requiring an interferometric measurement.  Here, emphasis has been placed on ``Gaussian-state'' because Pannu~\emph{et al}.'s recently proposed discrete-variable QI~\cite{Pannu2024} may avoid some of the problems that plague Gaussian-state QI.   
 
Pannu~\emph{et al.}\@ do \emph{not} assume that at most one photon is returned per transmission from the region being probed, hence avoiding the root cause of Lloyd's QI not outperforming its best classical competitor.  Their analysis is asymptotic, in that they pass to the limit $M\rightarrow\infty$ for the discrete-variable entangled state introduced in Lloyd's QI.  Doing so drives their receiver's false-alarm probability to zero, and makes it simple to prove that their receiver realizes 6\,dB quantum advantage in error-probability exponent in high-brightness noise, matching both Tan~\emph{et al}.'s performance in that regime and the Nair-Gu bound~\cite{Nair2020} on the attainable error-probability exponent of \emph{all} possible QI protocols.  Indeed, in this asymptotic regime, Pannu~\emph{et al}.'s QI matches the Nair-Gu bound at \emph{all} noise brightnesses.  However, contrary to their claim that Gaussian-state QI does not achieve Nair-Gu performance at low-noise brightness, it is easily shown~\cite{footnote1} that the system's error-probability exponent approaches the Nair-Gu bound, regardless of the noise brightness, as its signal brightness is decreased.   

It is obvious that \emph{requiring} infinite entangled-state dimensionality, i.e., $M\rightarrow \infty$, will put Pannu~\emph{et al}.\@ QI beyond the realm of practicality, and Ref.~\cite{Pannu2024} presents no finite-$M$ results for the error-probability exponent.  
Our paper remedies the preceding problem.  In particular, we introduce a new, more explicit, result for the joint density operator of the returned and retained radiation when the target is present.  Using this result we obtain accurate approximations to the single-shot false-alarm and detection probabilities for Pannu~\emph{et al}.'s receiver.  From those approximations we then get the finite-$M$ multi-shot likelihood-ratio test and use the Chernoff bound to derive its error-probability exponent.  We find that Pannu~\emph{et al.}'s QI has good and bad regimes, analogous to those of Lloyd's QI, and that only in the good regime---which requires  $M$ to exceed a threshold value---does it offer quantum advantage for the error-probability exponent.  

The rest of the paper is organized as follows.  In Sec.~\ref{Setup} we present the setup assumed by Pannu~\emph{et al}.\@ and summarize their key results.  In Sec.~\ref{DensityOperators} we derive the finite-$M$ joint density operators for the returned and retained radiation under target absence and presence, with the latter being much more amenable to finite-$M$ performance analysis than the form presented in Pannu~\emph{et al}.  In Sec.~\ref{single-shot} we find accurate approximations to the false-alarm and detection probabilities for a single finite-$M$ transmission, which we use in Sec.~\ref{multi-shot} to analyze multi-shot performance.  Section~\ref{discuss} concludes the main text with an appraisal of our results, including a comparison with Gaussian-state QI.  Appendix~\ref{AppendA} then contains derivation details for the target-present density operator, and Appendix~\ref{AppendB} shows that the first-order corrections to Sec.~\ref{single-shot}'s false-alarm and detection probability approximations lead to inconsequential changes in those results.  

\section{Pannu \emph{et al}.'s Quantum Illumination \label{Setup}}
Pannu~\emph{et al}.'s QI is an amalgam of Lloyd's QI and Tan~\emph{et al}.'s QI.  Thus, like Lloyd's QI transmitter, Pannu~\emph{et al}.'s QI transmitter prepares a sequence of $M$-dimensional, signal-idler ($S$-$I$) high-dimensional Bell states,  
\begin{equation}
|\psi\rangle_{SI} = \frac{1}{\sqrt{M}}\sum_{m=1}^M|\bem\rangle_S\,|\bem\rangle_I\,{}_I\langle\bem|\,{}_S\langle\bem|,
\label{BellState}
\end{equation}
where, for $K = S,I$, $|\bem\rangle_K$ denotes an $M$-mode Fock state for modes with annihilation operators $\{\hat{a}_{K_m}: 1\le m \le M\}$ containing 1 photon in mode $m$ and no photons in the remaining modes.   Again like Lloyd's QI, Pannu~\emph{et al}.'s QI transmits the signal modes from Eq.~(\ref{BellState}) into the region of interest, and retains the idler modes for a subsequent joint measurement with the radiation returned therefrom.  Using $\{\hat{a}_{R_m} : 1\le m \le M\}$ to denote the annihilation operators for those $M$ returned modes, Pannu~\emph{et al.}'s channel models for target absence and presence are those of Tan~\emph{et al}.   Specifically, under hypothesis $H_0$ (target absent) they use
\begin{equation}
\hat{a}_{R_m} = \hat{a}_{B_m},
\label{absence_model}
\end{equation}
where the $\{\hat{a}_{B_m} : 1\le m \le M\}$ are annihilation operators for background-noise modes that are in  independent, identically-distributed (iid) thermal states with average photon number $N_B$~\cite{footnote2}.  On the other hand, under hypothesis $H_1$ (target present), they use
\begin{equation}
\hat{a}_{R_m} = \sqrt{\kappa}\,\hat{a}_{S_m} +\sqrt{1-\kappa}\,\hat{a}_{B_m},
\label{presence_model}
\end{equation}
where $\kappa \ll 1$ is the roundtrip transmissivity to and from the weakly-reflecting target, and the iid background modes are now in thermal states with average photon number $N_B/(1-\kappa)$ to preclude the possibility of a passive signature of target presence.   

Pannu~\emph{et al}.'s insight---indeed their paper's great novelty---lies in its receiver's positive operator-valued measurement (POVM) for the returned and retained radiation from a single transmission.  On each transmission, Lloyd's QI receiver uses the POVM
\begin{equation}
\hat{\Pi}_k = \left\{\begin{array}{ll} 
|\psi\rangle_{RI}{}_{RI}\langle \psi|, & \mbox{for $k=1,$} \\[.05in]
\hat{I}_{RI}-\hat{\Pi}_1, & \mbox{for $k=0$},\end{array}\right.
\label{LloydPOVM}
\end{equation}
to decide $H_k$ was true, where 
\begin{equation}
|\psi\rangle_{RI} = \frac{1}{\sqrt{M}}\sum_{m=1}^M|\bem\rangle_R\,|\bem\rangle_I\,{}_I\langle\bem|\,{}_R\langle\bem|,
\end{equation}
and $\hat{I}_{RI}$ is the identity operator on the state space of the $\{\hat{a}_{R_m},\hat{a}_{I_m}\}$ modes.  Instead, for each transmission Pannu~\emph{et al}.'s receiver uses the POVM
 \begin{equation}
\hat{\Pi}_k = \left\{\begin{array}{ll} 
\sum_{\bfN}|\psi_\bfN\rangle_{RI}{}_{RI}\langle \psi_\bfN|, & \mbox{for $k=1,$} \\[.05in]
\hat{I}_{RI}-\hat{\Pi}_1, & \mbox{for $k=0$},\end{array}\right.
\label{PannuPOVM}
\end{equation}
where $\bfN \equiv (N_1,N_2,\ldots, N_M)$, $\sum_{\bfN} \equiv \prod_{m=1}^M\sum_{N_m=0}^\infty$, and 
\begin{equation}
|\psi_\bfN\rangle_{RI}\equiv \sum_{m=1}^M\sqrt{\frac{N_m+1}{|\bfN|+M}}\,|\bem+\bfN\rangle_R\,|\bem\rangle_I,
\end{equation}
with $|\bfN| \equiv \sum_{m=1}^MN_m$ and $|\bem+\bfN\rangle_R$ being the returned modes' state containing $N_m+1$ photons in the $\hat{a}_{R_m}$ mode and $N_{m'}$ photons in the $\{\hat{a}_{m'}:m'\neq m\}$ modes.  

Note that $|\psi_{\bf 0}\rangle_{RI} = |\psi\rangle_{RI}$, making Pannu~\emph{et al}.'s POVM a natural generalization of Lloyd's POVM to the Tan~\emph{et al}.\@ channel models, with their arbitrarily high numbers of photons in each returned mode.  So, with $\hat{\rho}^{(0)}_{RI}$ being the joint density operator for the $\{\hat{a}_{R_m},\hat{a}_{I_m}\}$ modes under hypothesis $H_0$, Pannu~\emph{et al}.\@ show that the $M$-mode, single-shot, false-alarm probability, $p_F \equiv {\rm Tr}(\hat{\Pi}_1\hat{\rho}^{(0)}_{RI})$, goes to zero as $M\rightarrow\infty$.    Similarly, they show that the $M$-mode, single-shot, detection probability, $p_D \equiv {\rm Tr}(\hat{\Pi}_1\hat{\rho}^{(1)}_{RI})$, obeys $\lim_{M\rightarrow\infty}p_D \ge \kappa/(N_B+1)$, where  $\hat{\rho}^{(1)}_{RI}$ is the joint density operator for the $\{\hat{a}_{R_m},\hat{a}_{I_m}\}$ modes under hypothesis $H_1$.  Because the false-alarm probability vanishes in this limit, they find that after $N_T\gg 1$ transmissions the multi-shot miss probability satisfies
\begin{equation}
\lim_{M\rightarrow\infty}P_M \le [1-\kappa/(N_B+1)]^{N_T} \approx e^{-\kappa N_T/(N_B+1)},
\end{equation}
where the approximation is valid because $\kappa \ll 1$.   For equally-likely target absence or presence, as we will assume in this paper, the $M\rightarrow \infty$ multi-shot error probability then obeys
\begin{equation}
\Pr(e) \le [1-\kappa/(N_B+1)]^{N_T} /2 \approx e^{-\kappa N_T/(N_B+1)}/2.
\label{PannuPr(e)}
\end{equation}

The Nair-Gu lower bound on QI error probability for equally-likely target absence or presence when $N_T$ signal photons are transmitted on average is~\cite{Nair2020}
\begin{equation}
\Pr(e)_{\rm LB} \ge [1-\kappa/(N_B+1)]^{N_T}/4 \approx e^{-\kappa N_T/(N_B+1)}/4,
\label{NairGuPr(e)}
\end{equation}
where the approximation uses $\kappa \ll 1$. This lower bound applies to optimum quantum reception for an \emph{arbitrary} choice of the signal-idler state, subject only to the constraint on the average transmitted photon number.  Comparing Eqs.~(\ref{PannuPr(e)}) and (\ref{NairGuPr(e)}) then shows that Pannu~\emph{et al}.'s receiver achieves the ultimate error-probability exponent, in the limit $M\rightarrow\infty$, for weakly-reflecting ($\kappa \ll 1$) targets at \emph{all} noise brightnesses.  Compared to its best classical competitor~\cite{Shapiro2009}, viz., a coherent-state (CS) system transmitting $N_T$ photons on average whose error-probability Chernoff bound is~\cite{Tan2008}
\begin{equation} 
\Pr(e)_{\rm CS} \le e^{-\kappa N_T(\sqrt{1+N_B}-\sqrt{N_B})^2}/2,
\label{CS-Pr(e)}
\end{equation}
Pannu~\emph{et al.}'s QI thus offers a 6\,dB quantum advantage in error-probability exponent when $N_B \gg 1$, no appreciable quantum advantage when $N_B \ll 1$, and 4.6\,dB quantum advantage at $N_B = 1$.  Tan~\emph{et al}.'s QI matches those behaviors because, as noted earlier, its error-probability exponent approaches the Nair-Gu bound in the limit of low signal brightness~\cite{footnote1}.  See Sec.~\ref{discuss} for a more detailed appraisal of Pannu~\emph{et al}.\@ QI versus Tan~\emph{et al}.\@ QI.

\section{Joint Density Operators \label{DensityOperators}}
The principal drawback of Ref.~\cite{Pannu2024}'s treatment of Pannu~\emph{et al}.'s QI is the absence of any finite-$M$ results for the error-probability exponent.  In this section we begin the task of obtaining such results by deriving a more useful form for the joint density operator for the single-shot returned and retained radiation when the target is present.  For completeness, however, we first present its target-absent counterpart, as that will be needed to evaluate the finite-$M$, single-shot, false-alarm probability.  From Eq.~(\ref{absence_model}) we immediately have that
\begin{equation}
\hat{\rho}^{(0)}_{RI} = \hat{\rho}^{(0)}_B\otimes\hat{\rho}_I,
\label{absence_densop}
\end{equation}
where
\begin{equation}
\hat{\rho}^{(0)}_B = \bigotimes_{m=1}^M\sum_{N_m=0}^\infty \frac{N_B^{N_m}}{(N_B+1)^{N_m+1}}\,|N_m\rangle_{R_m}\,{}_{R_m}\langle N_m|,
\end{equation}
with $|N_m\rangle_{R_m}$ being the $N_m$-photon state of the $\hat{a}_{R_m}$ mode, and
\begin{equation}
\hat{\rho}_I = {\rm Tr}_S(|\psi\rangle_{SI}\,{}_{SI}\langle \psi|) = \frac{1}{M}\sum_{m=1}^M |\bem\rangle_I\,{}_I\langle \bem|.
\end{equation}

To find the target-present joint density operator, $\hat{\rho}^{(1)}_{RI}$, we will use a characteristic-function approach to get its number-ket representation.  The anti-normally ordered characteristic function associated with $\hat{\rho}^{(1)}_{RI}$ is
\begin{equation}
\chi^{\rho^{(1)}_{RI}}_A(\bzeta_R,\bzeta_I) \equiv {\rm Tr}\!\left(\hat{\rho}^{(1)}_{RI}\,e^{-\bzeta_R^*\cdot\bhata_R-\bzeta_I^*\cdot\bhata_I}\,e^{\bzeta_R\cdot\bhata^\dagger_R+\bzeta_I\cdot\bhata^\dagger_I}\right),
\end{equation}
where, for $K=R,I$,  $\bzeta_K \equiv (\zeta_{K_1},\zeta_{K_2},\ldots, \zeta_{K_M})$ with $\{\zeta_{K_m}\}$ being complex valued, $\bhata_K \equiv (\hat{a}_{K_1},\hat{a}_{K_2},\ldots, \hat{a}_{K_M})$, and $\bhata^\dagger_K \equiv (\hat{a}^\dagger_{K_1},\hat{a}^\dagger_{K_2},\ldots, \hat{a}^\dagger_{K_M})$.  Using Eq.~(\ref{presence_model}), we can show that
\begin{equation}
\chi^{\rho^{(1)}_{RI}}_A(\bzeta_R,\bzeta_I) = \chi^{\rho_{SI}}_A(\sqrt{\kappa}\,\bzeta_R,\bzeta_I)\chi^{\rho^{(1)}_B}_A(\sqrt{1-\kappa}\,\bzeta_R),
\label{densop1}
\end{equation}
where $\chi^{\rho^{(1)}_B}_A(\bzeta_B) \equiv {\rm Tr}\!\left(\hat{\rho}^{(1)}_B e^{-\bzeta_B^*\cdot\bhata_B}\,e^{\bzeta_B\cdot\bhata^\dagger_B}\right)$ is the anti-normally ordered characteristic function associated with $\hat{\rho}^{(1)}_B$.  It is easily verified, using the assumed multi-mode thermal state for $\hat{\rho}^{(1)}_B$, that 
\begin{equation}
\chi^{\rho^{(1)}_B}_A(\sqrt{1-\kappa}\,\bzeta_R) = e^{-\bzeta_R^*\cdot\bzeta_R(N_B +1-\kappa)}.
\label{densop2}
\end{equation}

Next, to find $\chi^{\rho_{SI}}_A(\sqrt{\kappa}\,\bzeta_R,\bzeta_I)$, we first use the Baker-Campbell-Hausdorff theorem~\cite{Nielsen2000} to get
\begin{equation}
\chi^{\rho_{SI}}_A(\sqrt{\kappa}\,\bzeta_R,\bzeta_I) = e^{-\kappa\bzeta^*_R\cdot\bzeta_R -\bzeta^*_I\cdot\bzeta_I} \chi^{\rho_{SI}}_N(\sqrt{\kappa}\,\bzeta_R,\bzeta_I),
\label{densop3}
\end{equation}
where 
\begin{equation}
\chi^{\rho_{SI}}_N(\bzeta_S,\bzeta_I) \equiv {\rm Tr}\!\left(\hat{\rho}_{SI}\,e^{\bzeta_S\cdot\bhata^\dagger_S+\bzeta_I\cdot\bhata^\dagger_I}\,e^{-\bzeta^*_S\cdot\bhata_S-\bzeta^*_I\cdot\bhata_I}\right)
\label{chiNdefn}
\end{equation}
is the normally-ordered characteristic function associated with $\hat{\rho}_{SI}$. Expanding Eq.~(\ref{chiNdefn})'s exponential terms in Taylor series, and employing the result in Eq.~(\ref{densop3}), makes it easy to evaluate the latter equation.  Substituting the formula so obtained plus Eq.~(\ref{densop2}) into Eq.~(\ref{densop1}) gives us
\begin{align}
\chi^{\rho^{(1)}_{RI}}_A&(\bzeta_R,\bzeta_I) = e^{-\bzeta^*_R\cdot\bzeta_R(N_B+1)-\bzeta^*_I\cdot\bzeta_I} \nonumber \\[.05in]
&\times \left(1-\frac{\kappa \bzeta^*_R\cdot\bzeta_R}{M} - \frac{\bzeta^*_I\cdot\bzeta_I}{M} + \frac{\kappa |\bzeta_R\cdot\bzeta_I|^2}{M}\right).
\label{H1Chi_A}
\end{align}

Now it only remains for us to get the number-ket expansion of $\hat{\rho}^{(1)}_{SI}$ from its anti-normally ordered characteristic function via the operator-valued inverse Fourier transform,
\begin{align}
\hat{\rho}^{(1)}_{RI} &= \int\!\frac{{\rm d}^2\bzeta_R}{\pi^M}\int\!\frac{{\rm d}^2\bzeta_I}{\pi^M}\,\chi^{\rho^{(1)}_{RI}}_A(\bzeta_R,\bzeta_I) \nonumber \\[.05in]
&\times e^{-\bzeta_R\cdot\bhata^\dagger_R-\bzeta_I\cdot\bhata^\dagger_I} \,
e^{\bzeta^*_R\cdot\bhata_R+\bzeta^*_I\cdot\bhata_I},
\label{H1Chi_A_Inv}
\end{align} 
where $\int\!{\rm d}^2\bzeta_K/\pi^M\equiv \prod_{m=1}^M\int\!{\rm d}^2\zeta_{K_m}/\pi$ for $K = R, I$, and integrals without limits are from $-\infty$ to $\infty$ in all their dimensions.  

The rest of the derivation is rather involved, so it has been relegated Appendix~\ref{AppendA}.  The final expression is
\begin{widetext}
\begin{align}
\hat{\rho}^{(1)}_{RI} &= \frac{1}{M}\sum_{m=1}^M\sum_{\bfN}\!\left(\prod_{\ell =1}^M\frac{N_B^{N_\ell}}{(N_B+1)^{N_\ell+1}}\right)\!\left(1-\frac{\kappa}{N_B+1}+\frac{\kappa N_m \,u(N_m-1)}{N_B(N_B+1)}\right) |\bfN\rangle_R\,|\bem\rangle_I\,{}_I\langle \bem|\,{}_R\langle \bf N|\nonumber \\[.05in]
&+\frac{\kappa}{M}\sum_\bfN\sum_{\bfN'}\sum_{m=1}^M\sum_{m'=1\atop{m'\neq m}}^M\!\left(\prod_{\ell = 1\atop{\ell \neq m, m'}}^M\frac{N_B^{N_\ell}\,\delta_{N_\ell N'_\ell}}{(N_B+1)^{N_\ell+1}}\right)\frac{N_B^{N_{m'}+N'_m}}{(N_B+1)^{N_{m'}+N'_m+4}}\sqrt{(N_{m'}+1)(N'_m+1)}\nonumber \\[.05in]
&\times \left(\bigotimes_{\ell =1\atop{\ell \neq m,m'}}^M|N_\ell\rangle_{R_\ell}\right)\!|N'_m+1\rangle_{R_m}\,|N_{m'}\rangle_{R_{m'}}\,|\bem\rangle_I\,{}_I\langle \bemp|\,{}_{R_{m'}}\!\langle N_{m'}+1|\,{}_{R_m}\!\langle N'_m| \!\left(\bigotimes_{\ell =1\atop{\ell \neq m,m'}}^M{}_{R_\ell}\langle N'_\ell|\right),
\label{presence_densop}
\end{align}
\end{widetext}
where $u(\cdot)$ is the unit-step function and $\delta_{jk}$ is the Kronecker delta function.

\section{Single-Shot False-Alarm and Detection Probabilities \label{single-shot}}
The principal roadblock to obtaining finite-$M$ results for the single-shot false-alarm and detection probabilities is the $1/\sqrt{|\bfN|+M}$ factor in $|\psi_\bfN\rangle_{RI}$.  Because $M \gg 1$ is necessary to achieve an acceptably low error probability for the assumed weakly-reflecting target, especially in the case of high-brightness background noise, $|\bfN|+M$ will have high mean-to-standard-deviation ratios for all noise brightnesses under both the target absent and target present hypotheses.   Thus, in this section, we will replace $|\bfN|+M$ with its conditional means
in evaluating $p_F$ and $p_D$ from Eq.~(\ref{PannuPOVM})'s POVM and Eqs.~(\ref{absence_densop}) and (\ref{presence_densop})'s joint density operators for target absence and presence.  Appendix~\ref{AppendB} will show that the first-order corrections to this section's false-alarm and detection probability approximations are inconsequential.

To find our false-alarm probability approximation, we first rewrite $\hat{\rho}^{(0)}_{RI}$ as
\begin{equation}
\hat{\rho}^{(0)}_{RI} = \sum_{\tbfN}\Pr(\tbfN)\,|\tbfN\rangle_R\,{}_R\langle \tbfN|\otimes\frac{1}{M}\sum_{\tilde{m}=1}^M|\betm\rangle_I\,{}_I\langle \betm|,
\end{equation}
where $\Pr(\tbfN) \equiv \prod_{m=1}^MN_B^{\tilde{N}_m}/(N_B+1)^{\tilde{N}_m+1}$ and $|\tbfN\rangle_R \equiv \bigotimes_{m=1}^M|\tilde{N}_m\rangle_{R_m}$.  Then, we have that
\begin{align}
&p_F= \sum_\bfN{}_{RI}\langle \psi_\bfN|\hat{\rho}^{(0)}_{RI}|\psi_\bfN\rangle_{RI} \nonumber\\[.05in]
& =  \frac{1}{M}\sum_\bfN\sum_{\tilde{m}=1}^M\frac{N_{\tilde{m}}+1}{|\bfN|+M}
\sum_{\tbfN}\Pr(\tbfN)\,|{}_R\langle\betm+\bfN|\tbfN\rangle_R|^2.
\label{PfStep1}
\end{align}
Now, because 
\begin{equation}
|{}_R\langle\betm+\bfN|\tbfN\rangle_R|^2 = \delta_{\tilde{N}_{\tilde{m}}(N_{\tilde{m}}+1)}\prod_{m'=1\atop{m'\neq m}}^M \delta_{\tilde{N}_{m'}N_{m'}},
\end{equation}
Eq.~(\ref{PfStep1}) reduces to
\begin{align}
p_F &= \frac{1}{M}\sum_{\tilde{m}=1}^M\sum_{\tbfN}\Pr(\tbfN)\frac{\tilde{N}_{\tilde{m}}}{|\tbfN|+M-1} \\[.05in]
& \approx \frac{N_B}{M(N_B+1)-1},
\label{PfM}
\end{align}
where the approximation uses
\begin{align}
\frac{1}{|\tbfN|+M-1} &\approx \frac{1}{\sum_{\tbfN}\Pr(\tbfN)(|\tbfN|+M-1)} \\[.05in]
&= \frac{1}{M(N_B+1)-1}.
\label{rho0approx}
\end{align}   
Note that our $p_F$ approximation vanishes for $M\rightarrow\infty$, as found by Pannu~\emph{et al}.

Turning to the single-shot detection probability, we start from
\begin{align}
p_D& = \sum_\tbfN{}_{RI}\langle \psi_\tbfN|\hat{\rho}^{(1)}_{RI}|\psi_\tbfN\rangle_{RI}  \\[.05in]
& = \sum_\tbfN\sum_{m=1}^M\sum_{m'=1}^M\frac{\sqrt{(\tilde{N}_m+1)(\tilde{N}_{m'}+1)}}{|\tbfN|+M}\nonumber \\[.05in]
&\times {}_I\langle \bem|\,{}_R\langle \bem+\tbfN|\hat{\rho}^{(1)}_{RI}|\,|\bemp+\tbfN\rangle_R\,|\bemp\rangle_I.
\label{PdStep1}
\end{align}
To proceed further we will calculate the $m=m'$ and $m\neq m'$ components of Eq.~(\ref{PdStep1})
separately, using, respectively, the first line and second-plus-third lines of Eq.~(\ref{presence_densop}).  For the $m=m'$ terms we get
\begin{align}
p_D^{(m=m')} &= \frac{1}{M}\sum_\tbfN\sum_{m=1}^M\frac{\tilde{N}_m+1}{|\tbfN|+M}\nonumber \\[.05in] 
&\times \left(\prod_{\ell =1\atop{\ell \neq m}}^M\frac{N_B^{\tilde{N}_\ell}}{(N_B+1)^{\tilde{N}_\ell+1}}\right)\frac{N_B^{\tilde{N}_m+1}}{(N_B+1)^{\tilde{N}_m+2}} \nonumber \\[.05in]
&\times \left(1- \frac{\kappa}{N_B+1} +\frac{\kappa (\tilde{N}_m+1)}{N_B(N_B+1)}\right) \\[.05in]
&\approx \frac{1}{M}\sum_{m=1}^M\sum_{\tilde{N}_m = 0}^\infty \frac{\tilde{N}_m+1}{M(N_B+1)}\frac{N_B^{\tilde{N}_m+1}}{(N_B+1)^{\tilde{N}_m+2}} \nonumber \\[.05in]
&\times \left(1- \frac{\kappa}{N_B+1} +\frac{\kappa (\tilde{N}_m+1)}{N_B(N_B+1)}\right) \\[.05in]
&=\frac{\kappa +N_B}{M(N_B+1)},
\label{PdM}
\end{align}
where the approximation uses
\begin{align}
\frac{1}{|\tbfN|+M} &\approx \frac{1}{\sum_{\tbfN}\left(\prod_{m=1}^M \frac{N_B^{\tilde{N}_m}}{(N_B+1)^{\tilde{N}_m+1}}\right)\!(|\tbfN|+M)} \\[.05in]
&= \frac{1}{M(N_B+1)}.
\label{rho1approx}
\end{align}

For the $m\neq m'$ terms we get
\begin{align}
p&_D^{(m\neq m')} & \nonumber \\[.05in]
&= \frac{\kappa}{M}\sum_{m=1}^M\sum_{m'=1\atop{m'\neq m}}^M\sum_{\tilde{N}_m=0}^\infty\sum_{\tilde{N}_{m'}=0}^\infty \frac{N_B^{\tilde{N}_m+\tilde{N}_{m'}}}{(N_B+1)^{\tilde{N}_m+\tilde{N}_{m'}+4}} \nonumber \\[.05in]
&\times \frac{(\tilde{N}_m+1)(\tilde{N}_{m'}+1)}{|\tbfN|+M} \\[.05in]
&\approx \frac{\kappa (M-1)}{M(N_B+1)}.
\end{align}
where the approximation uses Eq.~(\ref{rho1approx}).

Putting the $m=m'$ and $m\neq m'$ results together we obtain the following approximation for the single-shot detection probability,
\begin{equation}
p_D \approx \frac{\kappa}{N_B+1} + \frac{N_B}{M(N_B+1)}.
\label{PdM}
\end{equation}

\section{Multi-Shot Likelihood Ratio and Error-Probability Exponent\label{multi-shot}}
In this section we first use Eqs.~(\ref{PfM}) and (\ref{PdM}) to determine the multi-shot likelihood-ratio test (LRT) for minimum error-probability choice between equally-likely target absence or presence based on the results of $N_T \gg 1$ Pannu~\emph{et al}.\@ single-shot POVMs.  From that LRT we then use the Chernoff bound to obtain its finite-$M$ error-probability exponent, from which we can identify the dimensionality threshold that must be exceeded for there to be any quantum advantage, and the minimum dimensionality required to be within 1\,dB of the Nair-Gu error-probability exponent.

Let $\{d_n : 1\le n\le N_T\}$ denote single-shot POVM results, i.e., $d_n = 1$ indicates a target-present decision on the $n$th transmission and $d_n=0$ denotes a target-absent decision on that transmission.  Conditioned on the true hypothesis, the $\{d_n\}$ are iid Bernoulli random variables with success probabilities $p_F$ for $H_0$ and $p_D$ for $H_1$.  The LRT we are seeking is therefore
\begin{equation}
\frac{\prod_{n=1}^{N_T}p_D^{d_n}[1-p_D]^{1-d_n}}{\prod_{n=1}^{N_T}p_F^{d_n}[1-p_F]^{1-d_n}} \begin{array}{c}\mbox{\scriptsize decide $H_1$}\\ \ge \\ < \\ \mbox{\scriptsize decide $H_0$}\end{array} 1,
\end{equation}
which can be rewritten as
\begin{equation}
\frac{\left(\begin{array}{c}N_T\\ D_{N_T}\end{array}\right)\!p_D^{D_{N_T}}[1-p_D]^{N_T-{D_{N_T}}}}{\left(\begin{array}{c}N_T\\ D_{N_T}\end{array}\right)\!p_F^{D_{N_T}}[1-p_F]^{N_T-{D_{N_T}}}} \begin{array}{c}\mbox{\scriptsize decide $H_1$}\\ \ge \\ < \\ \mbox{\scriptsize decide $H_0$}\end{array} 1,
\label{binomialLRT}
\end{equation}
with
\begin{equation} 
\left(\begin{array}{c}N_T\\ D_{N_T}\end{array}\right) \equiv \frac{N_T!}{D_{N_T}!(N_T-D_{N_T})!}
\end{equation} 
being the binomial coefficient and $D_{N_T}\equiv \sum_{n=1}^{N_T}d_n$. 

Equation~(\ref{binomialLRT}) shows that $D_{N_T}$ is a sufficient statistic for the minimum error-probability test, and the $D_{N_T}$-based LRT can be reduced to the simple threshold test, 
\begin{align}
D&_{N_T} \begin{array}{c}\mbox{\scriptsize decide $H_1$}\\ \ge \\ < \\ \mbox{\scriptsize decide $H_0$}\end{array} \nonumber \\[.05in]
&  \frac{N_T\ln\{[1-p_F]/[1-p_D]\}}{\ln\{p_D[1-p_F]/p_F[1-p_D]\}},
\end{align}
where we have assumed $M$ is large enough that $p_D > p_F$, i.e.,
\begin{equation}
M > M_0 \equiv\frac{\kappa +\sqrt{\kappa^2+4\kappa N_B(N_B+1)}}{2\kappa(N_B+1)}.
\end{equation}

The Chernoff bound we are seeking is
\begin{align}
\Pr&(e)\le \nonumber \\[.05in]
&\min_{0\le s\le 1}\sum_{n=0}^{N_T}\Pr(D_{N_T} = n\mid H_1)^s\Pr(D_{N_T} = n\mid H_0)^{1-s}/2.
\label{multi-shot_Chernoff}
\end{align}
For $M > M_0$, two simple calculations show that Eq.~(\ref{multi-shot_Chernoff})  reduces to 
\begin{equation}
\Pr(e) \le \min_{0\le s \le 1}[p_D^sp_F^{1-s} + (1-p_D)^s(1-p_F)^{1-s}]^{N_T}/2,
\end{equation}
and $s_o$, the minimizing $s$ value, is
\begin{align}
s&_o = \nonumber \\[.05in]
&\frac{\ln\{[p_F/(1-p_F)]\ln[(p_D/p_F)]/\ln[(1-p_F)/(1-p_D)]\}}{\ln[(1-p_D)p_F/(1-p_F)p_D]}.
\end{align}
To quantify Pannu \emph{et al}.\@ QI's approach to the Nair-Gu lower bound on QI's error-probability exponent, we introduce the penalty function $\mathcal{E}(\kappa, N_B, M)$ that satisfies
\begin{align}
\exp&\{-[\kappa N_T/(N_B+1)]\mathcal{E}(\kappa, N_B,M)\}  \nonumber \\[.05in]
&= [p_D^{s_o}p_F^{1-s_o} + (1-p_D)^{s_o}(1-p_F)^{1-s_o}]^{N_T},
\end{align}
and focus our attention on two special cases, $\kappa = 0.001$ with $N_B = 100$ and $\kappa = 0.001$ with $N_B=1$, as representatives of a weakly-reflecting embedded in either high-brightness or moderate-brightness noise~\cite{footnote3}.   

\begin{figure}[hbt]
    \centering
 \includegraphics[width=0.95\columnwidth]{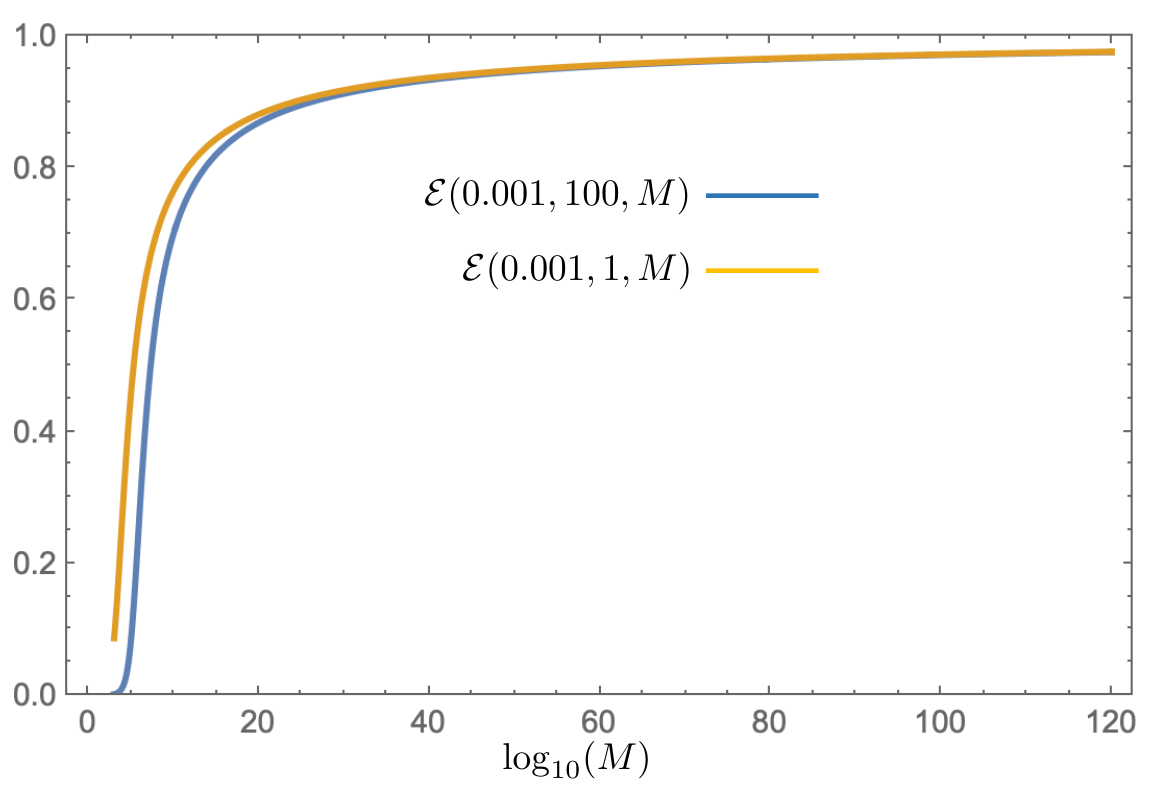}
    \caption{Plots of the penalty functions $\mathcal{E}(0.001, 1, M)$ (top curve) and $\mathcal{E}(0.001,100,M)$ (bottom curve) versus $\log_{10}(M)$. Note that for both curves the $M$ values shown here exceed the penalty functions' $M_0$ thresholds for $p_D > p_F$, viz., 22.6 for the moderate-brightness noise, and 31.5 for the high-brightness noise. \label{penalty_fig1}}    
\end{figure}

It is easily verified that $\lim_{M\rightarrow \infty}\mathcal{E}(\kappa, N_B, M) = 1$, so that our error-probability exponent matches the Nair-Gu bound at infinite $M$, as shown by Pannu~\emph{et al}.  But how high must $M$ be to approach that limit?   Figure~\ref{penalty_fig1} plots $\mathcal{E}(\kappa, N_B, M)$ versus $\log_{10}(M)$ for our representative cases, and Table~\ref{penalty_table} lists some key values therefrom.  For the high-brightness noise we see that: (1) 
below the $M = 7.34 \times 10^5 $ threshold, Pannu~\emph{et al}.\@ QI offers no quantum advantage in error-probability exponent;  and (2) $M = 2.27 \times 10^{13}$ is necessary for Pannu~\emph{et al}.\@ QI's error-probability exponent to be 1\,dB lower than the Nair-Gu bound's.  Similarly, for the moderate-brightness noise we find that:  (1) below the $M = 7.33 \times 10^3 $ threshold, Pannu~\emph{et al}.\@ QI offers no quantum advantage in error-probability exponent;  and (2) $M = 2.21 \times 10^{11}$ is necessary for Pannu~\emph{et al}.\@ QI's error-probability exponent to be 1\,dB lower than the Nair-Gu bound's.
\begin{table}[h]
\begin{center}
\begin{tabular}{|c|c|c|}\hline
$M$ & $\mathcal{E}(0.001, 100, M)$ & Comment\\ \hline \hline
$2.27 \times 10^{13}$ & $10^{-0.1}$ & 1\,dB off Nair-Gu bound \\ 
$7.34 \times 10^5$ & 0.25 & no quantum advantage \\ \hline\hline
$M$ & $\mathcal{E}(0.001, 1, M)$ & Comment \\ \hline \hline
$2.21 \times 10^{11}$ & $10^{-0.1}$ & 1\,dB off Nair-Gu bound \\ 
$7.33 \times 10^3$  & 0.25 & no quantum advantage \\ \hline\hline
\end{tabular}
\end{center}
\caption{$M$ values needed for representative values of $\mathcal{E}(0.001,100, M)$ and $\mathcal{E}(0.001,1,M)$. \label{penalty_table}}
\end{table}
  
\section{Conclusions and Discussion \label{discuss}}
Pannu~\emph{et al}.\@ launched a new paradigm for discrete-variable QI target detection.  They first combined the $M$ mode-pair signal-idler state from Lloyd's QI with the low-transmissivity channel models from Tan~\emph{et al.}'s Gaussian-state QI.  Then, they introduced a new single-shot POVM that enables the Nair-Gu bound on QI's error-probability exponent to be achieved at \emph{all} noise brightnesses in the limit $M\rightarrow \infty$.  Our work has established the finite-$M$ performance of Pannu~\emph{et al}.'s QI, showing that it has good and bad regimes---dictated by their entangled-states dimensionality---that are analogous to those of Lloyd~\emph{et al}.'s QI.   Furthermore, for any combination of roundtrip target transmissivity and background-noise brightness, it allows the entangled-state dimensionality needed to approach the Nair-Gu bound on QI's error-probability exponent to be quantified.

At this juncture, a comparison between finite-$M$ Pannu~\emph{et al}.\@ QI and Tan~\emph{et al}.\@ QI is warranted.  Both systems can match the Nair-Gu bound on target-detection error-probability exponent for a weakly-reflecting ($\kappa \ll 1$) target embedded in thermal noise.  Moreover, neither offers any appreciable quantum advantage for low-brightness ($N_B \ll 1$) background noise.  That said, the conditions required for each of these protocols to realize their respective quantum advantages are quite different.  

\begin{figure}[hbt]
    \centering
\includegraphics[width=0.975\columnwidth]{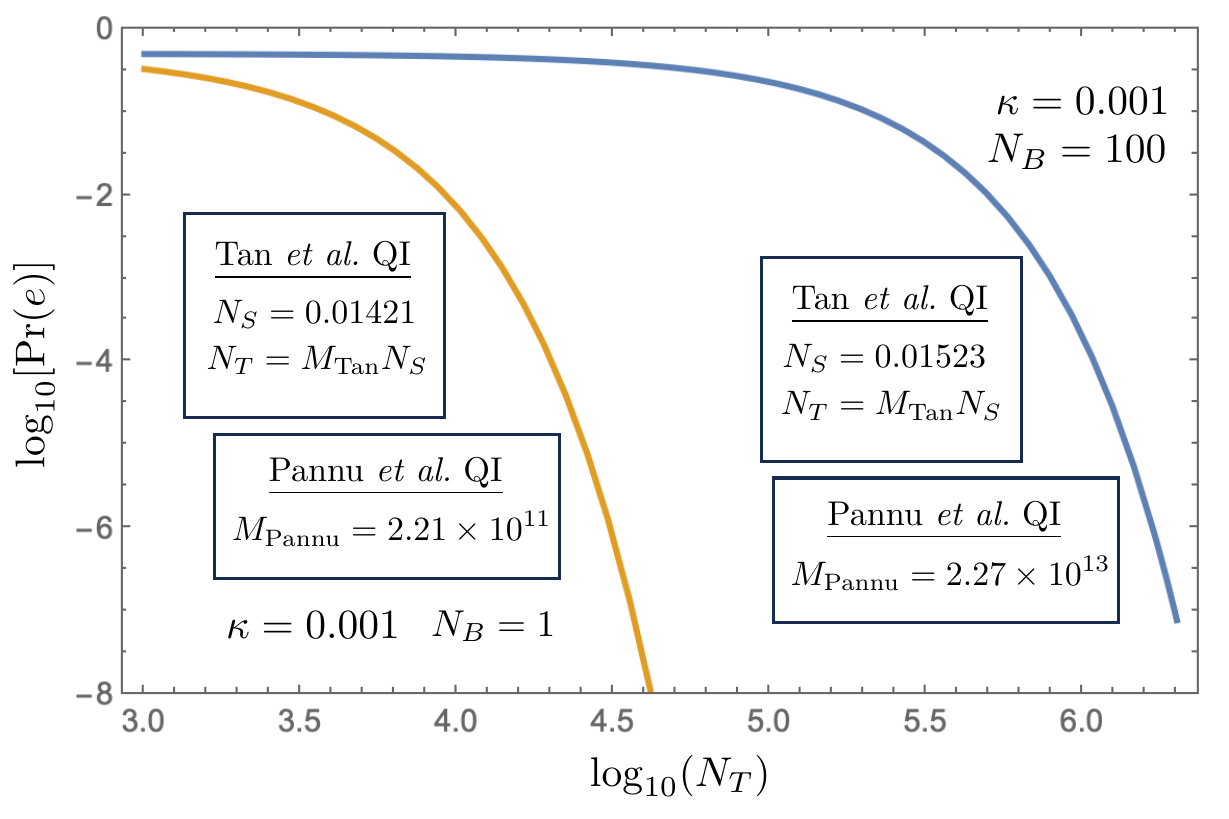}
    \caption{Log-log plots of the Chernoff bound, $\Pr(e)$, versus $N_T$ for Pannu~\emph{et al.}\@ QI and Tan~\emph{et al}.\@ QI when $\kappa = 0.001$ with $N_B = 100$ (top curve) and $\kappa = 0.001$ with $N _B = 1$ (bottom curve), with both systems are operating at 1\,dB below the Nair-Gu error-probability exponent.  In the high-brightness noise,  Pannu~\emph{et al.}\@ QI requires $M = 2.27 \times 10^{13}$ to reach this operating point, whereas in the moderate-brightness noise, $M= 2.21 \times 10^{11}$  suffices for this purpose.  In both brightnesses $N_T$ is the number of transmitted signal photons.  For Tan~\emph{et al}.\@ QI,  this operating point requires $N_S = 0.01523$ in the high-brightness noise and $N_B=0.01421$ in the moderate-brightness noise.  In both of these cases $N_T$ is now the average number of transmitted signal photons.    \label{Pe}}    
\end{figure}

Consider Pannu~\emph{et al}.\@ QI and Tan~\emph{et al.}\@ QI for our  $\kappa = 0.001$ with $N_B = 100$ and $\kappa = 0.001$ with $N_B =1$ examples, with both systems operating at error-probability exponents 1\,dB lower than the Nair-Gu bound's.  In the high-brightness noise, Pannu~\emph{et al}.'s QI requires $M = 2.27 \times 10^{13}$ to operate at 1\,dB below the Nair-Gu error-probability exponent, and achieves the Chernoff-bound performance shown in Fig.~\ref{Pe}, whereas in the moderate-brightness noise $M = 2.21 \times 10^{11}$ suffices for those purposes.  Tan~\emph{et al}.'s QI, on the other hand, requires the signal brightness to be $N_S = 0.01523$ to operate at 1\,dB below the Nair-Gu bound in the high-brightness noise, whereas $N_S =  0.01421$ suffices for that purpose in the moderate-brightness noise.  In both of those cases it achieves Fig.~\ref{Pe}'s Chernoff-bound performance, where $N_T =  MN_S$ is now the average number of transmitted signal photons.  

\begin{figure}[hbt]
    \centering
 \includegraphics[width=0.975\columnwidth]{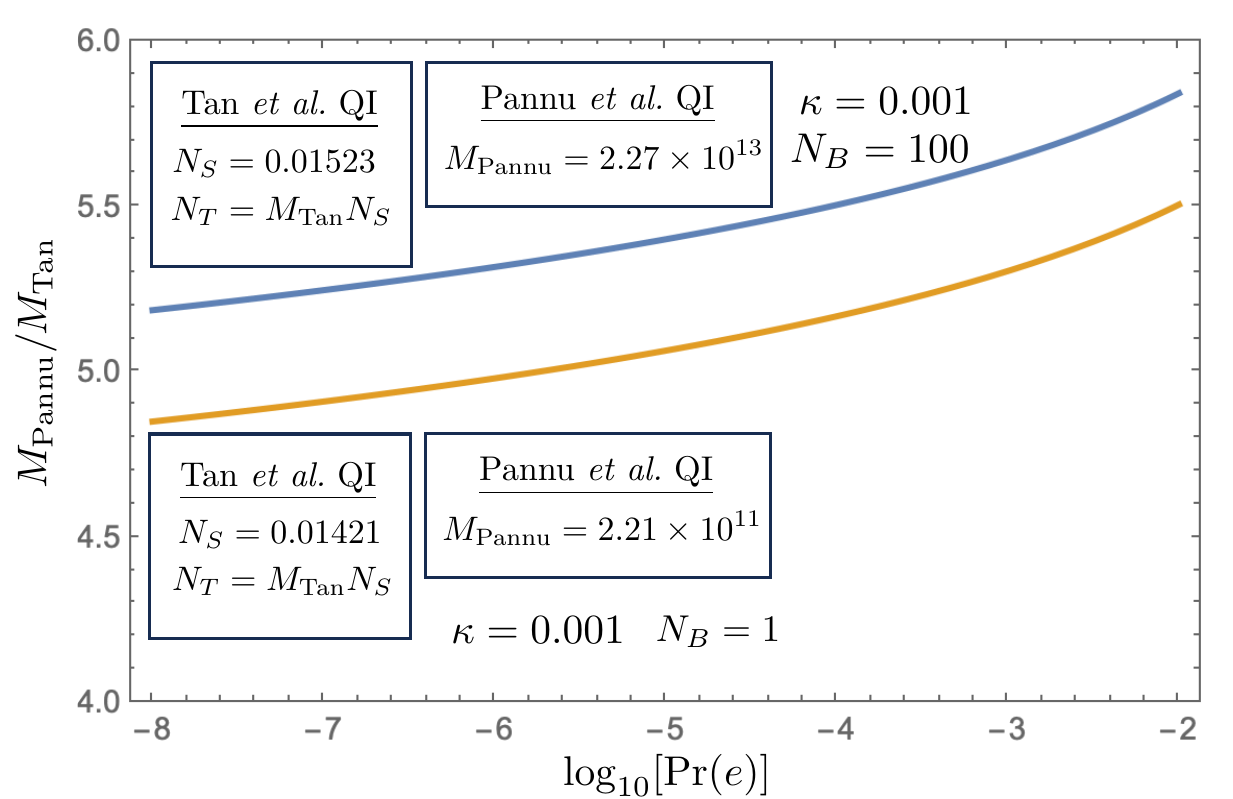}
    \caption{Log-log plots of $M_{\rm Pannu}/M_{\rm Tan}$ versus $\Pr(e)$ where $M_{\rm Pannu}$ and $M_{\rm Tan}$ are the entangled-state dimensionalities associated with the $\Pr(e)$ versus $N_T$ curves from Fig.~\ref{Pe}.    \label{Mcomp}}    
\end{figure}

Figure~\ref{Mcomp} compares the entanglement dimensionalities of the two QI systems for the parameters used in Fig.~\ref{Pe}.  Here we see that Pannu~\emph{et al}.\@ QI requires more than $10^5$ times the dimensionality that suffices for Tan~\emph{et al}.\@ QI in the high-brightness noise and more than $10^4$ times the dimensionality that Tan~\emph{et al}.\@ QI requires in the moderate-brightness noise.  

As if Pannu~\emph{et al}.\@ QI's requiring $10^4$-to-$10^5$ times the entangled-state dimensionality of Tan~\emph{et al}.\@ QI to achieve the same error probability were not bad enough, it incurs an even larger disadvantage when we look at time-bandwidth product.  In the scalar-wave, unresolved-target scenario that we are considering, only temporal degrees of freedom are available.  Hence, a single pulse of dimensionality $M$ must have a time duration $T$ and bandwidth $W$ satisfying $M = TW$.  Reference~\cite{Shapiro2020} explains that Tan~\emph{et al}.'s QI can carve $T = M/W$-s duration pulses from the signal and idler outputs of a continuous-wave-pumped parametric downconverter with $W$-Hz phase-matching bandwidth to obtain the $M$-dimensional entangled state it needs.  It is not clear how to generate the $M$-dimensional entangled-state pulses that Pannu~\emph{et al}.\@ QI needs, but a \emph{sequence} of $N_T > 10^4$ such pulses are needed for that QI system to realize the Chernoff-bound error probabilities $<10^{-3}$ from Fig.~\ref{Pe} in the moderate-brightness noise, and $N_T > 10^5$ pulses are required in the high-brightness noise.  Thus Pannu~\emph{et al}.\@ QI's \emph{total} required time-bandwidth product is at least $10^8$-to-$10^{10}$ times what suffices for Tan~\emph{et al}.'s QI to get to $\Pr(e) < 10^{-3}$ in our Fig.~\ref{Pe} examples.  

Our final points of comparison between these two QI protocols concern the four enormous hurdles, cited in Sec.~\ref{Intro}, that currently preclude finding a realistic target-detection use case for Tan~\emph{et al}.'s QI:  its need to interrogate one resolution bin at a time; its need for a quantum memory to store its high time-bandwidth product idler; its need for extremely high time-bandwidth product radiation to get an acceptably low error probability; and its need for an interferometric measurement.  How does Pannu~\emph{et al}.'s QI stack up against these hurdles?  There is at least some encouragement in this regard, but so far the overall prospects are pretty poor.  In particular, Pannu~\emph{et al}.'s QI still needs to interrogate one resolution bin at a time, and it still needs a quantum memory to store its high-dimensionality idler.  However, assuming Pannu~\emph{et al}.\@ QI's single-pulse idler dimensionality is the same as that of Tan~\emph{et al}.'s QI, its discrete-variable memory may be easier to implement than Tan~\emph{et al.}'s continuous-variable memory.  As we have seen, however,  Pannu~\emph{et al}.'s required single-pulse idler dimensionality is apt to be $10^4$-to-$10^5$ times that of Tan~\emph{et al}.'s QI.  Worse, Pannu~\emph{et al}.\@ QI's \emph{total} time-bandwidth product may have to be a factor of $10^8$-to-$10^{10}$ times that of Tan~\emph{et al}.\@ QI, owing to its need to transmit a long sequence of single-photon pulses.   A final positive note for Pannu~\emph{et al}.'s QI is that it does \emph{not} require an interferometric measurement.  This phase insensitivity follows from $\hat{\rho}^{(0)}_{RI}$ being diagonal in the number-ket basis, and the anti-normally ordered characteristic function associated with the joint density operator for the $\{e^{i\phi}\bhata_R, \bhata_I\}$ modes under hypotheses $H_1$ being given by Eq.~(\ref{H1Chi_A}) for all phase shifts $\phi$.

\acknowledgments
The author thanks A. Pannu, A. S. Helmy, and H. El Gamal for early access to their manuscript.  He also thanks R. Di Candia for alerting him to Ref.~\cite{DiCandia2021}. \\ 

\appendix

\section{Obtaining $\hat{\rho}^{(1)}_{RI}$ from $\chi^{\rho^{(1)}_{RI}}_A(\bzeta_R,\bzeta_I)$ \label{AppendA}} 
In this appendix we will derive the number-ket matrix elements of $\hat{\rho}^{(1)}_{RI}$ from Eq.~(\ref{presence_densop}).  We start by seeking the matrix elements for the first ($m=m'$) line of Eq.~(\ref{presence_densop}), beginning from
\begin{widetext}
\begin{align}
{}_I\langle \bem|\hat{\rho}^{(1)}_{RI}|\bem\rangle_{I} &= \int\!\frac{{\rm d}^2\bzeta_R}{\pi^M}\int\!\frac{{\rm d}^2\bzeta_I}{\pi^M}\,\chi_A^{\rho^{(1)}_{RI}}(\bzeta_R,\bzeta_I)e^{-\bzeta_R\cdot\bhata^\dagger_R} e^{\bzeta^*_R\cdot\bhata_R}{}_I\langle\bem|e^{-\bzeta_I\cdot\bhata^\dagger_I}\,e^{\bzeta^*_I\cdot\bhata_I}|\bem\rangle_{I} \\[.05in]
&= \int\!\frac{{\rm d}^2\bzeta_R}{\pi^M}\int\!\frac{{\rm d}^2\bzeta_I}{\pi^M}\,\chi_A^{\rho^{(1)}_{RI}}(\bzeta_R,\bzeta_I)e^{-\bzeta_R\cdot\bhata^\dagger_R}\,e^{\bzeta^*_R\cdot\bhata_R} (1-|\zeta_{I_m}|^2), 
\label{TaylorPd}
\end{align}
where Eq.~(\ref{TaylorPd}) follows from the Taylor-series expansions of $e^{-\zeta_{I_m}\hat{a}^\dagger_{I_m}}$ and $e^{\zeta^*_{I_m}\hat{a}_{I_m}}$.
 Now, substituting from Eq.~(\ref{H1Chi_A}) and recognizing that $e^{-\bzeta^*_I\cdot\bzeta_I}/\pi^M$ is the joint probability density function (jpdf) for the $\{\zeta_{I_m}\}$ to be a collection of $M$ iid circulo-complex Gaussian random variables with mean-squared magnitudes $\langle |\zeta_{I_m}|^2\rangle_{{I_m}} = 1$, we get
\begin{align}
{}_I\langle \bem|\hat{\rho}^{(1)}_{RI}&|\bem\rangle_{I} =  \int\!\frac{{\rm d}^2\bzeta_R}{\pi^M}\,
e^{-\bzeta^*_R\cdot\bzeta_R(N_B+1)}e^{-\bzeta_R\cdot\bhata^\dagger_R}e^{\bzeta^*_R\cdot\bhata_R}\!\left\langle\left[1-\frac{\bzeta^*_I\cdot\bzeta_I}{M}-\frac{\kappa\bzeta^*_R\cdot\bzeta_R}{M}+\frac{\kappa|\bzeta_R\cdot\bzeta_I|^2}{M}\right]
\frac{(1-|\zeta_{I_m}|^2)}{M}\right\rangle_I,
\end{align}
where $\langle\cdot\rangle_I$ denotes expected value over the jpdf $e^{-\bzeta^*_I\cdot\bzeta_I}/\pi^M$.  Complex-Gaussian moment factoring then reduces this expression to
\begin{align}
{}_I\langle \bem|\hat{\rho}^{(1)}_{RI}|\bem\rangle_{I} =  \int\!\frac{{\rm d}^2\bzeta_R}{\pi^M}\,
e^{-\bzeta^*_R\cdot\bzeta_R(N_B+1)}e^{-\bzeta_R\cdot\bhata^\dagger_R}e^{\bzeta^*_R\cdot\bhata_R}\frac{(1-\kappa |\zeta_{R_m}|^2)}{M}.
\end{align}
Next, using the Taylor-series expansions for $e^{-\bzeta\cdot\bhata^\dagger_R}$ and $e^{\bzeta^*_R\cdot\bhata_R}$, we find that
\begin{align}
{}_R\langle\bfN|\,{}_m\langle\bem|\hat{\rho}^{(1)}_{RI}|\bem\rangle_m\,|\bfN'\rangle_R &= 
\int\!\frac{{\rm d}^2\bzeta_R}{\pi^M}\,e^{-\bzeta^*_R\cdot\bzeta_R(N_B+1)}\!\left(\bigotimes_{\ell =1}^M\sum_{k_\ell=0}^{N_\ell} \sqrt{\frac{N_\ell!}{(N_\ell-k_\ell)!}}\frac{(-\zeta_{R_\ell})^{k_\ell}}{k_\ell!}{}_{R_\ell}\langle N_\ell-k_\ell|\right) \nonumber \\[.05in]
&\times \left(\bigotimes_{\ell' =1}^M\sum_{k'_{\ell'}=0}^{N'_{\ell'}} \sqrt{\frac{N'_{\ell'}!}{(N'_{\ell'}-k'_{\ell'})!}} \frac{\zeta^{*k'_{\ell'}}_{R_{\ell'}}}{k'_{\ell'}!} |N'_{\ell'}-k'_{\ell'}\rangle{R_{\ell'}}\right)
\frac{(1-\kappa |\zeta_{R_m}|^2)}{M}.
\label{diagStep1}
\end{align}

The bra-ket inner products in Eq.~(\ref{diagStep1}) vanish if $\ell \neq \ell'$, and, because $e^{-\bzeta^*_R\cdot\bzeta_R(N_B+1)}/[\pi/(N_B+1)]^M$ is the jpdf for the $\{\zeta_{R_m}\}$ to be a collection of $M$ iid circulo-complex Gaussian random variables with mean-squared magnitudes $\langle |\zeta_{R_m}|^2\rangle_{{R_m}} = 1/(N_B+1)$, complex-Gaussian moment factoring implies that only the $k_{\ell} = k'_{\ell}$ terms will survive in Eq.~(\ref{diagStep1}).  The inner products that remain after setting $\ell = \ell'$ and $k'_\ell = k_\ell$, viz., ${}_{R_\ell}\langle N_\ell-k_\ell|N'_\ell-k_\ell\rangle_{R_{\ell'}}$, then vanish unless $N'_\ell = N_{\ell}$, leading to
\begin{align}
{}_R\langle\bfN|\,{}_m\langle\bem|\hat{\rho}^{(1)}_{RI}|\bem\rangle_m\,|\bfN'\rangle_R = 
\int\!\frac{{\rm d}^2\bzeta_R}{\pi^M}\,e^{-\bzeta^*_R\cdot\bzeta_R(N_B+1)}\!\left(\bigotimes_{\ell =1}^M\delta_{N_\ell N'_\ell}\sum_{k_\ell=0}^{N_\ell} \left(\begin{array}{c}N_\ell \\ k_\ell\end{array}\right)\frac{(-|\zeta_{R_\ell}|^2)^{k_\ell}}{k_\ell!}\right)\frac{(1-\kappa |\zeta_{R_m}|^2)}{M}.
\label{diagStep2}
\end{align}
At this point we note that
\begin{equation}
\int\!\frac{{\rm d}^2\zeta_{R_\ell}}{\pi}\,e^{-|\zeta_{R_\ell}|^2(N_B+1)}\sum_{k_\ell=0}^{N_\ell} \left(\begin{array}{c}N_\ell \\ k_\ell\end{array}\right)\frac{(-|\zeta_{R_\ell}|^2)^{k_\ell}}{k_\ell!} = 
\frac{N_B^{N_\ell}}{(N_B+1)^{N_\ell +1}},
\label{diagStep3}
\end{equation}
and
\begin{equation}
\int\!\frac{{\rm d}^2\zeta_{R_m}}{\pi}\,e^{-|\zeta_{R_m}|^2(N_B+1)}\sum_{k_m=0}^{N_m} \left(\begin{array}{c}N_m \\ k_m\end{array}\right)\frac{(-|\zeta_{R_m}|^2)^{k_m+1}}{k_m!} = 
\frac{(N_m - N_B)N_B^{N_m-1}}{(N_B+1)^{N_m+2}},
\label{diagStep4}
\end{equation}
because the former is the number-ket matrix element ${}_{B_\ell}\langle N_\ell|\hat{\rho}^{(0)}_{B_\ell}|N_\ell\rangle_{B_\ell}$ of the operator-valued inverse Fourier transform of $\chi_A^{\rho^{(0)}_{B_\ell}}(\zeta_{B_\ell})$, and the latter is ${\rm d}({}_{B_\ell}\langle N_\ell|\hat{\rho}^{(0)}_{B_\ell}|N_\ell\rangle_{B_\ell})/{\rm d}N_B$. Substituting Eqs.~(\ref{diagStep3}) and (\ref{diagStep4}) into Eq.~(\ref{diagStep2}) then gives us the first line of Eq.~(\ref{presence_densop}).

Proceeding now to the $m\neq m'$ case, we assume $m\neq m'$ in all that follows, and start from
\begin{align}
{}_I\langle \bem|\hat{\rho}^{(1)}_{RI}|\bemp\rangle_{I} &= \int\!\frac{{\rm d}^2\bzeta_R}{\pi^M}\int\!\frac{{\rm d}^2\bzeta_I}{\pi^M}\,\chi_A^{\rho^{(1)}_{RI}}(\bzeta_R,\bzeta_I)e^{-\bzeta_R\cdot\bhata^\dagger_R} e^{\bzeta^*_R\cdot\bhata_R}{}_I\langle\bem|e^{-\bzeta_I\cdot\bhata^\dagger_I}e^{\bzeta^*_I\cdot\bhata_I}|\bemp\rangle_{I} \\[.05in]
&= \int\!\frac{{\rm d}^2\bzeta_R}{\pi^M}\int\!\frac{{\rm d}^2\bzeta_I}{\pi^M}\,\chi_A^{\rho^{(1)}_{RI}}(\bzeta_R,\bzeta_I) e^{-\bzeta_R\cdot\bhata^\dagger_R}e^{\bzeta^*_R\cdot\bhata_R} (-\zeta_{I_m}\zeta^*_{I_{m'}}).
\end{align}
Substituting in from Eq.~(\ref{H1Chi_A}), and using complex-Gaussian moment factoring, the preceding result reduces to 
\begin{align}
{}_I\langle \bem|\hat{\rho}^{(1)}_{RI}|\bemp\rangle_I &= \frac{\kappa}{M}\int\!\frac{{\rm d}^2\bzeta_R}{\pi^M}\,e^{-\bzeta^*_R\cdot\bzeta_R(N_B+1)}e^{-\bzeta_R\cdot\bhata^\dagger_R}  e^{-\bzeta^*_R\cdot\bhata_R}\langle\zeta_{R_{m'}}|\zeta_{I_{m'}}|^2\zeta^*_{R_m}|\zeta_{I_m}|^2\rangle_I \\[.05in]
&=\frac{\kappa}{M}\int\!\frac{{\rm d}^2\bzeta_R}{\pi^M}\,e^{-\bzeta^*_R\cdot\bzeta(N_B+1)}e^{-\bzeta_R\cdot\bhata^\dagger_R} e^{-\bzeta^*_R\cdot\bhata_R}(-\zeta_{R_{m'}}\zeta^*_{R_m}).
\label{offdiagStep1}
\end{align}
Using Taylor series expansions of $e^{-\bzeta_R\cdot\bhata^\dagger_R}$ and $e^{\bzeta^*_R\cdot\bhata_R}$ in Eq.~(\ref{offdiagStep1}), we next get

\begin{align}
{}_R\langle \bfN|\,{}_I\langle \bem|\hat{\rho}^{(1)}_{RI}|\bemp\rangle_I\,|\bfN'\rangle_R &= 
\frac{\kappa}{M}\int\!\frac{{\rm d}^2\bzeta_R}{\pi^M}e^{-\bzeta^*_R\cdot\bzeta_R(N_B+1)}
\left(\bigotimes_{\ell =1}^M\sum_{k_\ell = 0}^{N_\ell} \sqrt{\frac{N_\ell!}{(N_\ell-k_\ell)!}} \frac{(-\zeta_{R_\ell})^{k_\ell}(-\zeta_{R_{m'}})}{k_\ell!}\,{}_{R_\ell}\langle N_\ell - k_\ell |\right)
\nonumber \\[.05in]
& \times \left(\bigotimes_{\ell' =1}^M\sum_{k'_{\ell'} = 0}^{N'_{\ell'}} \sqrt{\frac{N'_{\ell'}!}{(N'_{\ell'}-{k'_{\ell'})!}}} \frac{\zeta^{*k'_{\ell'}}_{R_{\ell'}}\zeta^*_{R_m}}{k'_{\ell'}!} \,|N'_{\ell'} - k'_{\ell'} \rangle_{R_{\ell'}}\right).
\end{align}
Complex-Gaussian moment factoring now implies that for $\ell,\ell' \neq m,m'$ only the $\ell = \ell'$, $k_\ell = k'_\ell$ terms survive, in which case the bra-ket inner product shows that $N_\ell = N'_\ell$ also holds, leaving us with
\begin{align}
 {}_R&\langle \bfN|\,{}_I\langle \bem|\hat{\rho}^{(1)}_{RI}|\bemp\rangle_I\,|\bfN'\rangle_R = 
 \frac{\kappa}{M}\left(\prod_{\ell = 1\atop{\ell \neq m,m'}}^M\frac{N_B^{N_\ell}\delta_{N_\ell N_{\ell'}}}{(N_B+1)^{N_\ell +1}}\right)\int\!\frac{{\rm d}^2\zeta_{R_m}}{\pi}\int\!\frac{{\rm d}^2\zeta_{R_{m'}}}{\pi}\,e^{-(|\zeta_{R_m}|^2+|\zeta_{R_{m'}}|^2)(N_B+1)} \nonumber \\[.05in]
 &\times \left(\sum_{k_m=0}^{N_m}\sqrt{\frac{N_m!}{(N_m-k_m)!}}\frac{(-\zeta_{R_m})^{k_m}}{k_m!}\,{}_{R_m}\langle N_m-k_m|\right)\left(\sum_{k_{m'}=0}^{N_{m'}}\sqrt{\frac{N_{m'}!}{(N_{m'}-k_{m'})!}}\frac{(-\zeta_{R_{m'}})^{k_{m'}+1}}{k_{m'}!}\,{}_{R_{m'}}\langle N_{m'}-k_{m'}|\right) \nonumber \\[.05in]
 &\times \left(\sum_{k'_{m'}=0}^{N'_{m'}}  \sqrt{\frac{N'_{m'}!}{(N'_{m'}-k'_{m'})!}}\,\frac{\zeta_{R_{m'}}^{*k'_{m'}}}{k'_{m'}!}\, |N'_{m'}-k'_{m'}\rangle_{R_{m'}}\right)\left(\sum_{k'_m=0}^{N'_m} \sqrt{\frac{N'_m!}{(N'_m-k'_m)!}} \frac{\zeta_{R_m}^{*(k'_m+1)}}{k'_m!}|N'_m-k'_m\rangle_{R_m}\right).
 \end{align}
The only terms that survive the remaining complex-Gaussian moment factoring are those for which $k_m=k'_m+1$ and $k_{m'}+1 = k'_{m'}$, and these conditions imply that the only terms that survive after evaluation of the remaining bra-ket inner products are those for which $N_m=N'_m +1$ and $N_{m'}+1=N'_{m'}$, giving us
\begin{align}
{}_R&\langle \bfN|\,{}_I\langle \bem|\hat{\rho}^{(1)}_{RI}|\bemp\rangle_I\,|\bfN'\rangle_R = 
 \frac{\kappa}{M}\left(\prod_{\ell = 1\atop{\ell \neq m,m'}}^M\frac{N_B^{N_\ell}\delta_{N_\ell N_{\ell'}}}{(N_B+1)^{N_\ell +1}}\right)\int\!\frac{{\rm d}^2\zeta_{R_m}}{\pi}\int\!\frac{{\rm d}^2\zeta_{R_{m'}}}{\pi}\,e^{-(|\zeta_{R_m}|^2+|\zeta_{R_{m'}}|^2)(N_B+1)} \nonumber \\[.05in]
 &\times \left(\sum_{k'_m=0}^{N'_m}\left(\begin{array}{c}N'_m \\ k'_m\end{array}\right)\! \frac{\sqrt{N'_m+1}\,(-|\zeta_{R_m}|^2)^{k'_m+1}}{(k'_m+1)!}\right)\!\!
\left(\sum_{k_{m'}=0}^{N_{m'}}\left(\begin{array}{c}N_{m'} \\ k_{m'}\end{array}\right)\!\frac{\sqrt{N_{m'}+1}\,(-|\zeta_{R_{m'}}|^2)^{k_{m'}+1}}{(k_{m'}+1)!}\right)\!\delta_{N_m(N'_m+1)}\delta_{N'_m(N_{m'}+1)}.
\end{align}

To complete our derivation, we first perform the $\zeta_{R_m}$ and $\zeta_{R_{m'}}$ integrations, resulting in
\begin{align}
{}_R\langle \bfN|\,{}_I\langle \bem|\hat{\rho}^{(1)}_{RI}&|\bemp\rangle_I\,|\bfN'\rangle_R = 
\frac{\kappa}{M}\left(\prod_{\ell = 1\atop{\ell \neq m,m'}}^M\frac{N_B^{N_\ell}\delta_{N_\ell N_{\ell'}}}{(N_B+1)^{N_\ell +1}}\right)\!
\left(\sum_{k'_m=0}^{N'_m}\left(\begin{array}{c}N'_m \\ k'_m\end{array}\right)\!
\frac{(-1)^{k'_m+1}}{(N_B+1)^{k'_m+2}}\right) \nonumber \\[.05in]
&\times \left(\sum_{k_{m'}=0}^{N_{m'}}\left(\begin{array}{c}N_{m'} \\ k_{m'}\end{array}\right)\!\frac{(-1)^{k_{m'}+1}}{(N_B+1)^{k_{m'}+2}}\right)\! \sqrt{(N'_m+1)(N_{m'}+1)}\,\delta_{N_m(N'_m+1)}\delta_{N'_m(N_{m'}+1)},
\end{align}
and then do the binomial sums to get our final result,
\begin{align}
{}_R\langle \bfN|\,{}_I\langle \bem|\hat{\rho}^{(1)}_{RI}|\bemp\rangle_I\,|\bfN'\rangle_R &= 
\frac{\kappa}{M}\left(\prod_{\ell = 1\atop{\ell \neq m,m'}}^M\frac{N_B^{N_\ell}\delta_{N_\ell N_{\ell'}}}{(N_B+1)^{N_\ell +1}}\right)\!\frac{N_B^{N'_m+N_{m'}}\sqrt{(N'_m+1)(N_{m'}+1)}}{(N_B+1)^{N'_m+N_{m'}+4}}\,\delta_{N_m(N'_m+1)}\delta_{N'_m(N_{m'}+1)},
\end{align}
which verifies the second and third lines of Eq.~(\ref{presence_densop}).  
\end{widetext} 

\section{First-order corrections to Sec.~\ref{single-shot}'s $p_F$ and $p_D$ approximations \label{AppendB}}
In this appendix we assess the accuracies of our $p_F$ and $p_D$ approximations from Eqs.~(\ref{PfM}) and (\ref{PdM}).  Those formulas relied on approximating $1/(|\tbfN|+M-\delta_{k0})$, for $k=0$ and 1, respectively, by the $n=0$ term of
\begin{align}
\frac{1}{|\tbfN|+M-\delta_{k0}} &= \frac{1}{M(N_B+1)-\delta_{k0}} \nonumber \\[.05in]
&\times \left[\sum_{n=0}^\infty \left(\frac{-|\Delta \tbfN|}{M(N_B+1)-\delta_{k0}}\right)^n\right],
\label{Taylor}
\end{align}  
where $\Delta\tbfN \equiv (\Delta \tilde{N}_1,\Delta \tilde{N}_2,\ldots,\Delta \tilde{N}_M)$, with $\Delta \tilde{N}_m \equiv \tilde{N}_m - N_B$, being the fluctuating part of $\tbfN$, and $|\Delta\tbfN| \equiv \sum_{m=1}^M\Delta \tilde{N}_m$.  

The bracketed term in Eq.~(\ref{Taylor}) is a stochastic Taylor series.  Assuming $M\gg 1$, as will be the case for Pannu~\emph{et al.}\@ QI, the mean-to-standard-deviation ratio of $ x\equiv |\tbfN|+M -\delta_{k0}$ satisfies
\begin{equation}
\frac{\mathbb{E}(x)}{\sqrt{{\rm Var}(x)}} = \frac{M(N_B+1)-\delta_{k0}}{\sqrt{MN_B(N_B+1)}} \gg 1
\end{equation}
for all $N_B$ in both our $p_F$ and $p_D$ calculations.  Thus we will include the $n=1$ terms from Eq.~(\ref{Taylor}) and see how much that changes the results from Sec.~\ref{single-shot}.   

Including the $n=1$ term from Eq.~(\ref{Taylor}), we get that the false-alarm probability approximation becomes $N_B/[M(N_B+1)-1] -\Delta_F$ where
\begin{equation}
\Delta_F \equiv \frac{1}{M}\sum_{m=1}^M\sum_\tbfN \Pr(\tbfN)\frac{\tilde{N}_m|\Delta \tbfN|}{[M(N_B+1)-1]^2},
\end{equation}
with $\Pr(\tbfN) = \prod_{m=1}^M N_B^{N_m}/(N_B+1)^{N_m+1}$. It is now a straightforward calculation to show that
\begin{equation}
\Delta_F = \frac{N_B(N_B+1)}{[M(N_B+1)-1]^2}, 
\end{equation}
so that including the first-order correction we find that
\begin{equation}
p_F \approx\frac{N_B}{M(N_B+1)-1}\left(1-\frac{N_B+1}{M(N_B+1)-1}\right).
\end{equation}
This result shows that, for all $N_B$, the first-order correction has virtually no effect on the false-alarm probability approximation when $M\gg 1$.  

Our paper's last task is to obtain the detection probability approximation when we include the $n=1$ term from Eq.~(\ref{Taylor}).  Here we start from $p_D \approx (M\kappa + N_B)/M(N_B+1) - \Delta_D$, where 
\begin{widetext}
\begin{equation}
\Delta_D = \sum_\tbfN\sum_{m=1}^M\sum_{m'=1}^M\frac{|\Delta \tbfN|\sqrt{(\tilde{N}_m+1)(\tilde{N}_{m'}+1)}}{[M(N_B+1)]^2}\,{}_R\langle \bem+\tbfN|\hat{\rho}^{(1)}_{RI}|\bemp+\tbfN\rangle_R.
\end{equation}
As we did in Sec.~\ref{single-shot} for the $n=0$ detection-probability approximation, we will calculate the $m=m'$ and $m\neq m'$ terms in $\Delta_D$ separately.  For $\Delta^{(m=m')}_D$, we get
\begin{align}
\Delta^{(m=m')}_D &= \frac{1}{M}\sum_\tbfN\sum_{m=1}^M\frac{|\Delta\tbfN|(\tilde{N}_m+1)}{[M(N_B+1)]^2}\left(\prod_{\ell =1\atop{\ell \neq m}}^M\frac{N_B^{\tilde{N}_\ell}}{(N_B+1)^{\tilde{N}_\ell+1}}\right)\frac{N_B^{\tilde{N}_m+1}}{(N_B+1)^{\tilde{N}_m+2}} \left(1- \frac{\kappa}{N_B+1} +\frac{\kappa (\tilde{N}_m+1)}{N_B(N_B+1)}\right) \\[.05in]
&= \frac{N_B(3\kappa + N_B)}{[M(N_B+1)]^2} \ll p_D^{(m=m')},
\label{DeltaDeq}
\end{align}
for $M \gg N_B+1$, as will be the case in the quantum-advantage regime, cf.\@ Fig.~\ref{penalty_fig1}.  Finally, for $m\neq m'$ we have
\begin{align}
\Delta_D^{(m\neq m')} &= \frac{\kappa}{M}\sum_{m=1}^M\sum_{m'=1\atop{m'\neq m}}^M\sum_{\tilde{N}_m=0}^\infty\sum_{\tilde{N}_{m'}=0}^\infty \frac{N_B^{\tilde{N}_m+\tilde{N}_{m'}}}{(N_B+1)^{\tilde{N}_m+\tilde{N}_{m'}+4}} \frac{|\Delta\tbfN|(\tilde{N}_m+1)(\tilde{N}_{m'}+1)}{[M(N_B+1)]^2} \\[.05in]
&= \frac{\kappa (M-1)N_B^2}{[M(N_B+1)]^2} \ll p_D^{(m\neq m')}, 
\label{DeltaDneq}
\end{align}
for $M \gg N_B+1$, as will be the case in the quantum-advantage regime, cf.\@ Fig.~\ref{penalty_fig1}.  Putting together Eqs.~(\ref{DeltaDeq}) and (\ref{DeltaDneq}) we have that the first-order correction to our $p_D$ approximation from Sec.~
\ref{single-shot} has essentially no effect.   
\end{widetext}


\begin{thebibliography}{99}
\bibitem{footnote3}In the high-brightness case, $N_B = 100$, the Nair-Gu bound gives 6\,dB quantum advantage, and in the moderate-brightness case, $N_B = 1$, that bound gives 4.6\,dB quantum advantage.  For $N_B = 0.01$ low-brightness noise, the Nair-Gu bound gives 0.82\,dB quantum advantage, thus a system operating 1\,dB away from that bound offers no quantum advantage.
\end{thebibliography}
\end{document}